\documentclass[preprint,showpacs,showkeys,aps]{revtex4}

\def\q1{{q^{-1}}}
\def\qq1{{q-q^{-1}}}

\begin{document}

\title{Classical $q$-deformed dynamics}
\author{A. Lavagno$^{1,2}$, A.M. Scarfone$^{1,3}$ and P.Narayana Swamy $^4$}
\address{$^1$Dipartimento di Fisica, Politecnico di Torino, Italy}
\address{$^2$Istituto Nazionale di Fisica Nucleare (INFN), Sezione di Torino, Italy}
\address{$^3$Istituto Nazionale di Fisica della Materia (CNR--INFM), Sezione di Torino, Italy}
\address{$^4$Southern Illinois University, Edwardsville, IL 62026, USA}

\begin {abstract}
On the basis of the quantum $q$-oscillator algebra in the
framework of quantum groups and non-commutative
$q$-differential calculus, we investigate a possible
$q$-deformation of the classical Poisson bracket in order
to extend a generalized $q$-deformed dynamics in the
classical regime. In this framework, classical $q$-deformed
kinetic equations, Kramers and Fokker-Planck equations, are
also studied.
\end {abstract}
\maketitle


\section{Introduction}
The study of quantum algebras and quantum groups has
attracted a lot of interest in the last few years,  and
stimulated intensive research in several physical fields in
view of a broad spectrum of applications, ranging from
cosmic strings and black holes to the fractional quantum
Hall effect and high-$T_c$ superconductors \cite{wil}.\\
From the seminal work of Biedenharn \cite{bie} and
Macfarlane \cite{mac} it is clear that the $q$-calculus,
originally introduced at the beginning of last century by
Jackson \cite{jack} in the study of the basic
hypergeometric function, plays a central role in the
representation of the quantum groups \cite{exton}. In fact
it has been shown that it is possible to obtain a
``coordinate" realization of the Fock space of the
$q$-oscillators by using the deformed Jackson derivative
(JD) or the so-called $q$-derivative operator
\cite{flo,cele1,fink}.

In this paper we want to introduce a $q$-deformation of the PB
($q$-PB) in order to define a generalized $q$-deformed dynamics in
a $q$-commutative phase-space. For this purpose we begin with the
observation that the creation and annihilation operators in the
quantum $q$-deformed SU$_q$(2) algebra corresponds  classically to
$q$-commuting coordinates in a $q$-phase space and that the
commutation relation between the standard quantum operators
corresponds classically to the Poisson bracket (PB).\\ The
motivation for our goal lies in the fact that a full understanding
of the physical origin of $q$-deformation in classical physics is
still lacking because it is not clear if there exists a classical
counterpart to the $q$-deformed quantum mechanics inspired by the
study of quantum groups. The problem of a possible $q$-deformation
of classical mechanics was dealt with in Ref. \cite{Klimek} where
a $q$-PB was obtained starting from a point of view different from
the one adopted in this paper.  In order to introduce the
classical correspondence of the quantum $q$-oscillator,  we shall
follow the main approach based on the following idea. The
(undeformed) quantum commutation relations are invariant under the
action of SU(2) and, as a consequence, the $q$-deformed
commutation relations are invariant under the action of SU$_q$(2).
Analogously, since the (undeformed) PB is invariant under the
action of the symplectic group Sp(1), we have to require that
$q$-PB must satisfy invariance under the action of the
$q$-deformed symplectic group Sp$_{q}$(1).

\section{Non-commutative differential calculus}
Since the creation and annihilation operators in the
quantum $q$-deformed SU$_q$(2) algebra correspond
classically to non-commuting coordinates in a
$q$-phase-space, in this section we introduce the
$q$-deformed plane which is generated by the
non-commutative elements $\hat x$ and $\hat p$ fulfilling
the relation \cite{Reshetikhin}
\begin{equation}
\hat p\,\hat x=q\,\hat x\,\hat p \ ,
\label{qplane}
\end{equation}
which is invariant under GL$_q$(2) transformations.
Henceforward, for simplicity, we shall limit ourselves to
consider the two-dimensional case.\\ From Eq.(\ref{qplane})
the $q$-calculus on the $q$-plane can be obtained formally
through the introduction of the $q$-derivatives
$\hat\partial_x$ and $\hat\partial_p$ \cite{Wess}
\begin{eqnarray}
&&\hat\partial_p\,\hat p=\hat\partial_x\,\hat x=1 \ ,\\
&&\hat\partial_p\,\hat x=\hat\partial_x\,\hat p=0 \ .
\end{eqnarray}
They fulfill the $q$-Leibniz rule
\begin{eqnarray}
&&\hat\partial_p\,\hat p=1+q^2\,\hat
p\,\hat\partial_p+(q^2-1)\,\hat x\,\hat\partial_x \ ,\label{pp}\\
&&\hat\partial_p\,\hat x=q\,\hat x\,\hat\partial_p\ ,\label{px}\\
&&\hat\partial_x\,\hat p=q\,\hat p\,\hat\partial_x\ ,\label{xp}\\
&&\hat\partial_x\,\hat x=1+q^2\,\hat x\,\hat\partial_x \
,\label{xx}
\end{eqnarray}
together with the $q$-commutative derivative
\begin{equation}
\hat\partial_p\,\hat\partial_x=q^{-1}\,\hat\partial_x\,\hat\partial_p \ .
\end{equation}
It is easy to see that in the $q\rightarrow 1$ limit one recovers
the ordinary commutative calculus. Let us outline the asymmetric
mixed derivative relations,  Eq.(\ref{pp}) and Eq.(\ref{xx}),  in
$\hat x$ and in $\hat p$. These properties arise directly from the
non-commutative structure of the $q$-plane defined in
Eq.(\ref{qplane}).

We recall now that the most general function on the $q$-plane can
be expressed as a polynomial in the $q$-variables $\hat x$ and
$\hat p$
\begin{equation}
f(\hat x,\,\hat p)=\sum_{i,j}c_{ij}\,\hat x^i\,\hat p^j \ ,
\end{equation}
where we have assumed the $\hat x$-$\hat p$ ordering prescription
(it can  always be accomplished by means of Eq.(\ref{qplane}).
Thus, taking into account Eqs.(\ref{pp})-(\ref{xx}), we obtain the
action of the $q$-derivatives on the monomials
\begin{eqnarray}
&&\hat\partial_x(\hat x^n\,\hat p^m)=[n]_q\,\hat x^{n-1}\,\hat p^m
\ ,\label{dxmon}\\
&&\hat\partial_p(\hat x^n\,\hat p^m)=[m]_q\,q^n\,\hat x^n\,\hat
p^{m-1} \ ,\label{dpmon}
\end{eqnarray}
where we have introduced the  $q$-basic number
\begin{equation}
[n]_q=\frac{q^{2n}-1}{q^2-1} \ .
\end{equation}

A realization of the above $q$-algebra and its $q$-calculus can be
accomplished by the replacements \cite{Ubriaco}
\begin{eqnarray}
&&\hat x\rightarrow x \ ,\label{x}\\
&&\hat p\rightarrow p\,D_x \ ,\label{p}\\
&&\hat\partial_x\rightarrow{\cal D}_x \ ,\label{dx}\\
&&\hat\partial_p\rightarrow{\cal D}_p\,D_x \ ,\label{dp}
\end{eqnarray}
where
\begin{eqnarray}
&&D_x=q^{x\,\partial_x} \ ,\\
&&D_x f(x,\,p)=f(q\,x,\,p) \ ,
\end{eqnarray}
is the dilatation operator along the $x$ direction (reducing to
the identity operator for $q\rightarrow 1$), whereas
\begin{eqnarray}
&&{\cal D}_x=\frac{q^{2\,x\,\partial_x}-1}{(q^2-1)\,x} \
,\\&&{\cal D}_p=\frac{q^{2\,p\,\partial_p}-1}{(q^2-1)\,p}\
,
\end{eqnarray}
are the JD with respect to $x$ and $p$. Their action on an
arbitrary function $f(x,\,p)$ is
\begin{eqnarray}
&&{\cal
D}_x\,f(x,\,p)=\frac{f(q^2\,x,\,p)-f(x,\,p)}{(q^2-1)\,x} \
,\\ &&{\cal
D}_p\,f(x,\,p)=\frac{f(x,\,q^2\,p)-f(x,\,p)}{(q^2-1)\,p} \
.
\end{eqnarray}
Therefore, as a consequence of the non-commutative structure of
the $q$-plane, in this realization the $\hat x$ coordinate becomes
a $c$-number and its derivative is the JD whereas the $\hat p$
coordinate and its derivative are realized in terms of the
dilatation operator and JD.

\section{$q$-Poisson bracket and $q$-symplectic group}

With the formulation of the $q$-differential calculus, we are now
able to introduce a $q$-PB.  Since the undeformed PB is invariant
under the action of the undeformed symplectic group Sp(1), we will
assume as previously stated, as a fundamental point, that the
$q$-PB must satisfy the invariance property under the action of
the $q$-deformed symplectic group Sp$_q$(1) with the same  value
of the deformed parameter $q$ used in the construction of the
quantum plane.

Let us start by recalling the classical definition of a 2-Poisson
manifold, which is a two dimensional Euclidean space $I\!\!R^2$
generated by the position and momentum variables $x\equiv x^1$ and
$p\equiv x^2$ and equipped with a PB. By introducing $f(x,\,p)$
and $g(x,\,p)$,  two arbitrary smooth functions, the PB is defined
as \cite{gold}
\begin{equation}
\Big\{f,\,g\Big\}=\partial_xf\,\partial_pg-
\partial_pf\partial_xg \
.\label{poisson}
\end{equation}
Eq. (\ref{poisson}) can be expressed in a compact form
\begin{equation}
\Big\{f,\,g\Big\}=\partial_{i}f\,J^{ij}\,\partial_{j}g \
,\label{poisson1}
\end{equation}
where $J^{ij}$ are the entries of the unitary symplectic
matrix $J$ given by

\begin{equation}
J=\left(
\begin{array}{cc}
0&1\\ -1&0
\end{array}
\right) \ .\label{j}
\end{equation}

Remarkably, Eq.(\ref{poisson1}) does not change under the
action of a symplectic transformation Sp(1) on the
phase-space. As  is well known Eq. (\ref{poisson1}) can
also be expressed as
\begin{equation}
\Big\{f,\,g\Big\}=\{x^i,\,x^j\}\,\partial_{i}f\,\partial_{j}g
\ ,
\end{equation}
so that, if we know the PB between the generators $x^i$ we
can compute the PB between any pair of functions $f$ and
$g$.

By requiring that the $q$-PB must be invariant  under the action
of the $q$-symplectic group Sp$_{q}$(1), we are lead to introduce
the following $q$-deformed PB between the $q$-generators
$\hat{x}^i$ \cite{slw}
\begin{equation}
\Big\{\hat x^i,\,\hat x^j\Big\}_q=\hat\partial_x\,\hat
x^i\,\hat\partial_p\,\hat x^j-q^2\,\hat\partial_p\,\hat
x^i\,\hat\partial_x\,\hat x^j \ .\label{qpoisson1}
\end{equation}
It is easy to verify the following fundamental relations
\begin{eqnarray}
&&\Big\{\hat x,\,\hat x\Big\}_q=\Big\{\hat p,\,\hat p\Big\}_q=0 \ ,\\
&&\Big\{\hat x,\,\hat p\Big\}_q=1 \ ,\label{pxp}\\
&&\Big\{\hat p,\,\hat x\Big\}_q=-q^2 \ ,\label{ppx}
\end{eqnarray}
which coincide with the one obtained in Ref. \cite{Klimek}. In
particular, from Eqs. (\ref{pxp}) and (\ref{ppx}) it follows that
the $q$-PB is not antisymmetric. A similar behavior appears also
in the quantum $q$-oscillator theory \cite{bie,mac}.

By means of Eqs.(\ref{x})-(\ref{dp}), a realization of our
generalized $q$-PB can be written as
\begin{equation}
\!\!\!\!\!\!\!\!\!\!\!\!\!\Big\{f,\,g\Big\}_q={\cal D}_x\,f(x,\,p\,D_x)\,{\cal
D}_p\,g(q\,x,\,p\, D_x)-q^2\,{\cal D}_p\,f(q\,x,\,p\,
D_x)\,{\cal D}_x\,g(x,\,p\, D_x) \label{qpoisson2}
\end{equation}
where $f$ and $g$ are identified with $x$ or $p$,
respectively.

\section{$q$-deformed kinetic equations}

On the basis of the above classical $q$-deformed theory, we
shall now derive the corresponding classical kinetic
equations based on the $q$-calculus. Starting from the
realization of the $q$-algebra, defined in Eqs.
(\ref{x})-(\ref{dp}), we are able to write the Kramers
equation corresponding to the equation of motion for the
distribution function $f(x,\,p;\,t)$, in position and
momentum space, describing the motion of particles of mass
$m$ in an external field $F(x)$ \cite{risken}. In the
one-dimensional case it can be generalized as follows

\begin{eqnarray}
\frac{\partial f(x,\,p;\,t)}{\partial t}=\Big
\{-&&\frac{p}{m} {\cal D}_x D_x- {\cal D}_p D_x \,
[J_1^q(p\,D_x) +F(x)]+ \nonumber \\
&&J_2^q \,({\cal D}_p\,D_x) \,({\cal D}_p\,D_x) \Big \}\,
f(x,\,p;\,t) \; ,
\end{eqnarray}
where $J_1^q(p\,D_x)$ and $J_2^q$ are the drift and
diffusion coefficients, respectively. Specifying the action
of the dilatation operator $D_x$ along the $x$ direction,
the above Kramers equation can be written as
\begin{eqnarray}
\frac{\partial f(x,\,p;\,t)}{\partial t}=-&&\frac{p}{m}
\,{\cal D}_x \,f(q\,x,\,p;\,t)
-{\cal D}_p \,[J_1^q(p\,D_x)+F(q\,x)] f(q\,x,\,p;\,t)]+ \nonumber \\
&& J_2^q \,{\cal D}_p^2  \,f(q^2\,x,\,p;\,t) \; ,
\end{eqnarray}
where ${\cal D}_p^2$ means the double application of the JD in the
momentum space. Without any external force, for a homogeneous
system undergoing  a constant diffusion, the above generalized
Kramers equation reduces to the following $q$-deformed
Fokker-Planck equation

\begin{equation}
\frac{\partial f(p;\,t)}{\partial t}={\cal D}_p
\Big[-J_1^q(p) + J_2^q \, {\cal D}_p \Big ]\,  f(p;\,t) \;
.
\end{equation}
If we postulate a generalized Brownian motion in a $q$-deformed
classical dynamics by  the following definition of the drift and
diffusion coefficients

\begin{equation}
J_1^q(p)= -\, \gamma \, p \;\; \left (\frac{q^2
\,D_p+1}{2}\right ) \; , \ \ \ \ \ J_2^q=\gamma \, m \, k T
\; ,
\end{equation}
where $\gamma$ is the friction constant, $T$ is the temperature of
the system and $D_p$ is the dilatation operator in the momentum
space, the stationary solution $f_{\rm st}(p)$ of the above
Fokker-Planck equation can be obtained as a solution of the
following stationary $q$-differential equation
\begin{equation}
{\cal D}_p f(p)=-\frac{p}{2\,m\,kT} [q^2\, f(qp)+f(p)] \; .
\end{equation}
It is easy to show that the solution of the above equation can be
written as
\begin{equation}
f_{\rm st}(p)=E_q\left[-\frac{p^2}{2m \, kT}\right ] \; ,
\end{equation}
where $E_q[x]$ is the $q$-deformed exponential function,
well-known in  $q$-calculus \cite{exton}, defined in terms
of the series

\begin{equation}
E_q[x]=\sum_{k=0}^{\infty} \frac{x^k}{[k]_q!}
\end{equation}
where $[k]_q!$ is the $q$-basic factorial \cite{exton}
defined as $[k]_q!=[k]_q\,[k-1]_q\cdots[1]_q$.

\section{Conclusions}

We have shown that $q$-calculus can play a crucial role in
the formulation of a generalized $q$-classical theory,
defined by means of the introduction of a $q$-PB. In
analogy with quantum group invariance properties of the
quantum $q$-oscillator theory, the $q$-PB has been defined
by assuming the invariance under the action of Sp$_q$(1)
group with its derivatives acting on the $q$-deformed
non-commutative plane invariant under Gl$_q$(2)
transformations. Therefore such a classical $q$-deformation
theory can be seen as the analogue of $q$-oscillator
deformation in the quantum theory. In this framework, we
have studied the classical $q$-deformed kinetic equations,
Kramers and Fokker-Planck equations and we have found, as a
stationary solution, the well-known $q$-deformed
exponential function defined in terms of a series. This
opens the possibility of introducing a classical
counterpart of the quantum $q$-deformations and we expect
that such a classical $q$-deformed dynamics can be very
relevant in several physical applications.


\end{document}